\documentclass[aps,prd,superscriptaddress,nofootinbib,preprint]{revtex4}

\usepackage{ulem}
\usepackage{color}
\usepackage{graphicx}
\usepackage{multirow}
\usepackage{amsmath,amssymb,amsthm,amsxtra,overpic,bbm,bm,float
,epsfig}
\newcommand{\tabincell}[2]{\begin{tabular}{@{}#1@{}}#2\end{tabular}}
\usepackage{color}

\fontsize{3}{5}\selectfont

\begin{document}

\begin{center}
{\Large\bf Probing Direct and Indirect Unitarity Violation in Future Accelerator Neutrino Facilities}
\end{center}

\vspace{0.3cm}

\begin{center}
{\bf Jian Tang} \footnote{E-mail: tangjian5@mail.sysu.edu.cn},
~{\bf Yibing Zhang}
\footnote{E-mail: zhangyb27@mail2.sysu.edu.cn} \\
{\sl School of Physics, Sun Yat-Sen University, Guangzhou, China}
\end{center}

\begin{center}
{\bf Yu-Feng Li}
\footnote{E-mail: liyufeng@ihep.ac.cn}  \\
{\sl Institute of High Energy Physics, Chinese Academy of Sciences, and School of Physical Sciences, University of Chinese Academy of Sciences, Beijing 100049, China}
\end{center}

\setcounter{footnote}{0}

\vspace{2.5cm}

\section*{\large Abstract}

The possible existence of light and heavy sterile neutrinos may give rise to \textit{direct} and \textit{indirect} unitarity violation
of the $3\times3$ lepton mixing matrix respectively. In the current work we study the potential of future accelerator neutrino facilities in probing
the unitarity violation effects. Taking DUNE, T2HK and a low-energy Neutrino Factory (LENF) as working examples of future accelerator neutrino
facilities, we study the distinct effects of direct and indirect unitarity violation on the discovery reach of the leptonic CP violation and
precision measurements of $\theta_{23}$ in the three neutrino framework. In addition, constraints on the additional mixing parameters
of direct and indirect unitarity violation are also discussed. Finally, we stress that the combination of experiments with different oscillation channels, different neutrino
beams and different detector techniques will be an effective solution to the parameter degeneracy problem and give the robust measurement of leptonic CP violation even
if the direct and indirect unitarity violation are taken into account.

\newpage

\section{Introduction}

The experimental establishment of neutrino oscillations~\cite{Kajita:2016cak,McDonald:2016ixn} requires non-zero neutrino masses and non-trival lepton mixing.
With current data from solar, atmospheric, reactor and accelerator neutrino oscillation experiments~\cite{Olive:2016xmw}, we can define a
standard three neutrino oscillation framework with two independent mass-squared differences, where the solar (SOL) and atmospheric (ATM) mass-squared differences
are given as $\Delta{m}^2_{\rm SOL} = \Delta{m}^2_{21} \simeq 7.5 \times 10^{-5} \, \rm{eV}^2$ and
$\Delta{m}^2_{\rm ATM} = |\Delta{m}^2_{31}| \simeq |\Delta{m}^2_{32}| \simeq 2.4 \times 10^{-3} \, \rm{eV}^2$ respectively.
The sign of $\Delta{m}^2_{31}$ is still unknown and referred as the neutrino mass ordering problem. Regarding the lepton flavor
mixing, it is described by a $3\times3$ unitary matrix, the Pontecorvo-Maki-Nakagawa-Sakata (PMNS) matrix~\cite{Pontecorvo:1957cp,Maki:1962mu},
which in the standard parametrization~\cite{Olive:2016xmw} can be described in terms of three mixing angles (i.e., $\theta_{12}$, $\theta_{23}$, $\theta_{13}$)
and one CP-violating phase (i.e., $\delta$). At present, we have acquired good knowledge for the three mixing angles, but only limited information on
the CP-violating phase. Therefore, measurements of the neutrino mass ordering and the CP-violating phase are the primary goal
for future neutrino oscillation experiments. In addition, precision measurements of neutrino oscillations also give us opportunities to
test the completeness of the three neutrino oscillation framework and to search for new physics beyond the Standard Model.

Additional sterile neutrino states, no matter whether they are heavy or light, will result in the unitarity violation of the $3\times3$ PMNS matrix if
they are mixed with the three active neutrinos. The anomalies from short baseline neutrino oscillation experiments have provided interesting hints
for the existence of light sterile neutrinos~\cite{sterile,Gariazzo:2017fdh,Gariazzo:2015rra,Giunti:2013aea,Kopp:2013vaa}.
On the other hand, sterile neutrinos at the keV scale are excellent candidates for the warm dark matter~\cite{Adhikari:2016bei,Kusenko:2009up,Boyarsky:2009ix,Araki:2011zg}.
From the theoretical point of view of the neutrino mass generation, sterile neutrinos are regarded as natural
ingredients of the canonical seesaw models~\cite{Minkowski:1977sc,Yanagida:1979ss,Gell-Mann:1979ss,Glashow:1979ss,Mohapatra:1979ia}.
However, the number of sterile neutrinos is arbitrary and the corresponding mass scale
could range from the sub-eV scale to the scale of the Grand Unification Theories (GUTs) of $10^{15}$ GeV.

For low energy neutrino oscillation phenomena, unitarity violation~\cite{Antusch:2006vwa,Xing:2012kh,Qian:2013ora,Luo:2014fia,Antusch:2014woa,Escrihuela:2015wra,Li:2015oal,Parke:2015goa,Fong:2016yyh,Blennow:2016jkn}
 is a generic consequence of sterile neutrinos and therefore can be probed in
the precision measurements of the neutrino oscillation experiments. When sterile neutrinos are heavy and kinematically forbidden in the neutrino
production and detection processes, they only have indirect effects on the $3\times3$ PMNS matrix, which is defined
as indirect unitarity violation (IUV)~\cite{Xing:2012kh,Luo:2014fia,Li:2015oal,Fong:2016yyh,Blennow:2016jkn}.
In contrast, when the sterile neutrinos are light and participate in the neutrino production, oscillation and detection
processes, their direct effects on the the $3\times3$ PMNS matrix will be very different from the heavy case. We define this scenario as the direct
unitarity violation (DUV)~\cite{Xing:2012kh,Luo:2014fia,Li:2015oal,Fong:2016yyh,Blennow:2016jkn}.
Strictly speaking, the boundary of the mass scale for the IUV and DUV scenarios is process dependent~\cite{Li:2015oal,Fong:2016yyh,Blennow:2016jkn},
but roughly it is located at the electro-weak interaction scale.

In a previous work done by one of the present authors~\cite{Li:2015oal}, we have derived the neutrino oscillation probabilities in matter with a constant density
considering direct and indirect unitary violation in the lepton mixing matrix. Analytical approximation and numerical calculation for the distinct
effects of the DUV and IUV are discussed. In the current work we are going to study the potential to probe the DUV and IUV effects in future accelerator neutrino facilities. 
Taking the Deep Underground Neutrino Experiment (DUNE)~\cite{Acciarri:2015uup},
the Tokai-to-Hyperkamiokande experiment (T2HK)~\cite{Abe:2014oxa} and a low energy Neutrino Factory (LENF)~\cite{FernandezMartinez:2010zza} as the representative examples,
we employ the GLoBES software package~\cite{Huber:2004ka,Huber:2007ji} to simulate the aforementioned experiments, and discuss the distinct features of the DUV and IUV
in the measurements of the lepton CP violating phase and the $\theta_{23}$ octant. We also discuss
the experimental sensitivity to constrain the new mixing parameters induced by the DUV and IUV effects.

This work is organized as follows. In section 2 we briefly discuss the theoretical framework of neutrino oscillations in the presence of the DUV and IUV.
In section 3, we describe the main setups of the considered experiments and their realization in GLoBES. The simulation results and physics interpretations
are presented in detail in section 4. Finally we summarize in section 5.

\section{Theoretical framework}

\begin{figure}[!t]
\begin{center}
\includegraphics[scale=0.35]{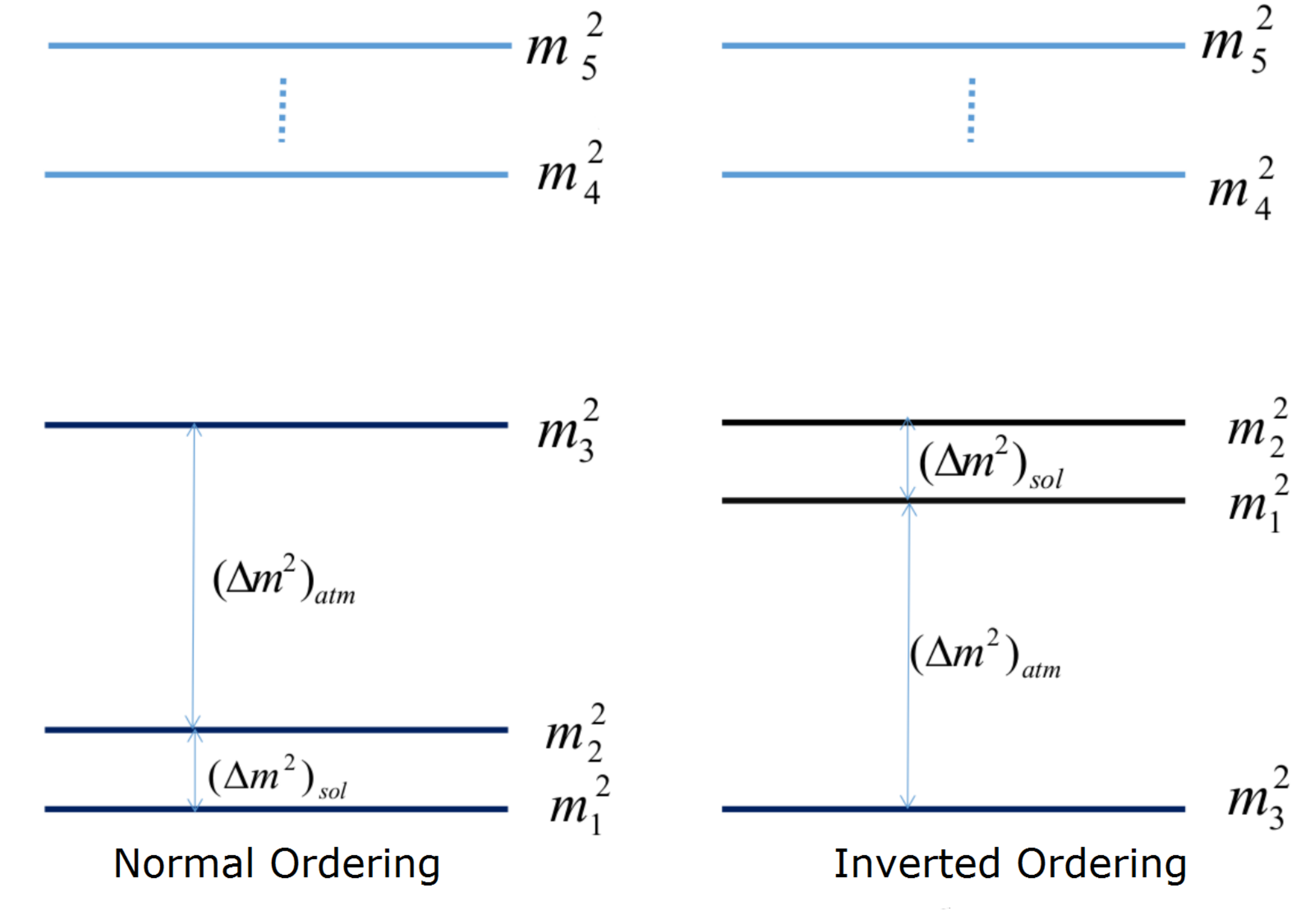}
\caption{The mass ordering schemes of the phenomenological $3$+$\mathbbm{1}$+$\mathbf{1}$ model in the current study. }
\label{fig:mass ordering}
\end{center}
\end{figure}

In this work, we consider the simplest ($3$+$\mathbbm{1}$+$\mathbf{1}$) scenario with two additional species of sterile neutrinos, of which one is the light sterile neutrino and the other one is the heavy sterile one~\footnote{If we consider the canonical seesaw realization of the neutrino mass generation~\cite{Minkowski:1977sc,Yanagida:1979ss,Gell-Mann:1979ss,Glashow:1979ss,Mohapatra:1979ia}, there will be at least two heavy sterile neutrinos.
Here we take the phenomenological $3$+$\mathbbm{1}$+$\mathbf{1}$ model to make the discussion as simple as possible. Effects of the scenarios with arbitrary numbers of light and heavy sterile neutrinos
are the same as this simplest case.}.
Fig.~\ref{fig:mass ordering} shows two possible cases for the mass ordering of three active neutrinos,
where we consider both the light and heavy sterile neutrinos. Note that in the current work we only discuss the case of the normal mass ordering.
the DUV and IUV effects for the case of the inverted mass ordering are similar to the normal mass ordering case.

In the ($3$+$\mathbbm{1}$+$\mathbf{1}$) scenario, the full neutrino mixing is parametrized by a $5 \times 5$ unitary matrix ${\cal U}$,
which can be decomposed as \cite{Xing2012,Dev:2012bd}
\begin{eqnarray}
{\cal U} \; = \; \left ( \begin{matrix} {\bm 1} & {\bm 0} \cr {\bm 0} & U^{}_{0} \cr \end{matrix} \right ) \left ( \begin{matrix} A & R \cr S & B \cr \end{matrix} \right ) \left ( \begin{matrix} V^{}_{0} & {\bm 0} \cr {\bm 0} & {\bm 1} \cr \end{matrix} \right ) \; = \; \left ( \begin{matrix} A V^{}_{0} & R \cr U^{}_{0} S V^{}_{0} & U^{}_{0} B \cr \end{matrix} \right ) \; ,
\end{eqnarray}
where  $U^{}_{0}$ and $V^{}_{0}$ are the unitary matrices while $A$, $B$, $R$ and $S$ are not, $\bm{1}$ and $\bm{0}$ stand for the identity and zero matrices respectively.
The detailed discussions of these matrices can be found in Ref.~\cite{Xing2012}.
The matrix $V^{}_{0}$ in Eq.~(1) can be parametrized using the standard parametrization as~\cite{Olive:2016xmw}
\begin{eqnarray}
V^{}_{0} \; = \; \left ( \begin{matrix} c^{}_{12} c^{}_{13} & s^{}_{12} c^{}_{13} & s^{}_{13} e^{-i\delta}_{} \cr -s^{}_{12} c^{}_{23} - c^{}_{12} s^{}_{23} s^{}_{13} e^{i\delta}_{} & c^{}_{12} c^{}_{23} - s^{}_{12} s^{}_{23} s^{}_{13} e^{i\delta}_{} & s^{}_{23} c^{}_{13} \cr s^{}_{12} s^{}_{23} - c^{}_{12} c^{}_{23} s^{}_{13} e^{i\delta}_{} & - c^{}_{12} s^{}_{23} - s^{}_{12} c^{}_{23} s^{}_{13} e^{i\delta}_{} & c^{}_{23} c^{}_{13} \end{matrix} \right ) \; .
\end{eqnarray}
To the order of $s^{2}_{ij}$ ($i = 1, 2, 3\; {\rm and}\; j = 4, 5$), $A$ and $R$ can be approximately written as~\cite{Xing2012}
\begin{eqnarray}
A  \simeq  {\bm 1} - \left( \begin{matrix} \displaystyle\frac{1}{2} \left( s^2_{14} + s^2_{15} \right) & 0 & 0 \cr \hat{s}^{}_{14} \hat{s}^*_{24} + \hat{s}^{}_{15} \hat{s}^*_{25} & \displaystyle\frac{1}{2} \left( s^2_{24} + s^2_{25} \right) & 0 \cr \hat{s}^{}_{14} \hat{s}^*_{34} + \hat{s}^{}_{15} \hat{s}^*_{35} & \hat{s}^{}_{24} \hat{s}^*_{34} + \hat{s}^{}_{25} \hat{s}^*_{35} & \displaystyle\frac{1}{2} \left( s^2_{34} + s^2_{35} \right) \end{matrix} \right) \;,\quad\quad 
R  \simeq  \left( \begin{matrix} \hat{s}^*_{14} & \hat{s}^*_{15} \cr \hat{s}^*_{24} & \hat{s}^*_{25} \cr \hat{s}^*_{34} & \hat{s}^*_{35} \end{matrix} \right) \; ,
\end{eqnarray}
where $s^{}_{ij} \equiv \sin\theta^{}_{ij}$, $c^{}_{ij} \equiv \cos\theta^{}_{ij}$ and $\hat{s}^{}_{ij} \equiv e^{i \delta^{}_{ij}}_{} \sin\theta^{}_{ij}$ with $\theta^{}_{ij}$ and $\delta^{}_{ij}$ are the rotation and phase angles, respectively. Note that although there are 6 additional mixing angles and 6 phases in $A$ and $R$, only 6 mixing angles and 4 independent phases will affect the low energy neutrino oscillation experiments. The other two independent phases are of the Majorana type and do not appear in the neutrino oscillations~\cite{Bilenky:1980cx}. These parameters can be divided into two separated groups:
\begin{itemize}
\item  ($\theta^{}_{14}$, $\theta^{}_{24}$, $\theta^{}_{34}$) and ($\delta^{}_{14}$, $\delta^{}_{24}$ $\delta^{}_{34}$) are parameters of the DUV describing the mixing between three active neutrinos and the light sterile neutrino. Since the light sterile neutrino can be produced and detected in the low energy experiments, these parameters will also appear in the neutrino oscillation probabilities as those standard mixing parameters.
\item  ($\theta^{}_{15}$, $\theta^{}_{25}$, $\theta^{}_{35}$) and ($\delta^{}_{15}$, $\delta^{}_{25}$, $\delta^{}_{35}$) are the IUV parameters that describe the mixing between three active neutrinos and the heavy sterile neutrino. The IUV affects the processes of neutrino production and detection, and therefore induce the zero-distance effects~\cite{Langacker:1988up,Kopp:2007ne,Agarwalla:2014bsa} in the neutrino oscillations.
\end{itemize}
To discuss the low energy phenomena of neutrino oscillations, the heavy sterile neutrino is not accessible and needs to be integrated out.
Therefore we are left with a $ 4\times 4$ non-unitary mixing matrix describing the mixing of three active neutrinos and a light sterile neutrino:
\begin{eqnarray}
\left ( \begin{matrix} \nu^{}_{e} \cr \nu^{}_{\mu} \cr \nu^{}_{\tau} \cr \nu^{}_{s} \end{matrix} \right ) \; = \; U \left ( \begin{matrix} \nu^{}_{1} \cr \nu^{}_{2} \cr \nu^{}_{3} \cr \nu^{}_{4} \end{matrix} \right ) \; = \; \left ( \begin{matrix} U^{}_{e1} & U^{}_{e2} & U^{}_{e3} & U^{}_{e4} \cr U^{}_{\mu 1} & U^{}_{\mu 2} & U^{}_{\mu 3} & U^{}_{\mu 4} \cr U^{}_{\tau 1} & U^{}_{\tau 2} & U^{}_{\tau 3} & U^{}_{\tau 4} \cr U^{}_{s 1} & U^{}_{s 2} & U^{}_{s 3} & U^{}_{s 4}\end{matrix} \right ) \; \left ( \begin{matrix} \nu^{}_{1} \cr \nu^{}_{2} \cr \nu^{}_{3} \cr \nu^{}_{4} \end{matrix} \right ) \; ,
\end{eqnarray}
where $U$ is just the $4 \times 4$ left-up truncated sub-matrix of the full $5 \times 5$ mixing matrix $\cal U$.
One can use $U$ to calculate the oscillation probabilities in vacuum and in matter for the ($3$+$\mathbbm{1}$+$\mathbf{1}$) mixing scenario,
where both the analytical approximation and the numerical calculation can be found in Ref.~\cite{Li:2015oal} by turning off the mixing parameters of
the second heavy sterile neutrino.
In the current study, we use the full numerical method to calculate the neutrino oscillation probabilities in matter.

\section{Experimental setups}

In this section we first describe the simulation details for DUNE~\cite{Acciarri:2015uup}, T2HK~\cite{Abe:2014oxa} and the LENF~\cite{FernandezMartinez:2010zza},
which are shown in Tab.~\ref{tab:glbtable} with the neutrino beam sources, detector setups and systematics
and the running information provided.
\begin{table}
\footnotesize
\centering
\begin{tabular}{|c|c|c|c|c|}
\hline
Experiments &T2HK & DUNE & LENF\\
\hline
Neutrino beam& \tabincell{c}{1.3 MW power, \\$2.7\times10^{21}$ POT/yr}&\tabincell{c}{1.07 MW power,\\$1.47\times10^{21}$ POT/yr}& \tabincell{c}{stored muons: \\$10.66 \times 10^{20}$/yr,\\ $E_{\mu}$ = 5$ \;\rm GeV$}  \\
\hline
\tabincell{c}{Detector\\(fiducial mass)} &\tabincell{c}{WC (190 kton/1 tank)}& LArTPC (40 kton) & TASD (20 kton)      \\
\hline
Baseline & 295 km & 1300 km & 1200 km \\
\hline
Energy resolution &$8.5\%/\sqrt{E}$ & Migration matrices & $15\%/\sqrt{E}$\\
\hline
Runtime& \tabincell{c}{${\nu}$ 1.5 yrs + $\bar{\nu}$ 4.5 yrs with 1 tank\\${\nu}$ 1 yrs + $\bar{\nu}$ 3 yrs with 2 tanks}& $\nu$ 3.5 yrs + $\bar{\nu}$ 3.5 yrs &  $\nu$ 4 yrs + $\bar{\nu}$ 4 yrs \\
\hline
Energy range & 0.4 GeV to 1.2 GeV & 0.5 GeV to 10 GeV &  0.5 GeV to 5 GeV \\
\hline
\tabincell{c}{Normalization\\error on signal} & \tabincell{c}{$2.5\%$ (all channels)} &$1\%$ (all channels)& $2\%$ (all channels)\\
\hline
\tabincell{c}{Normalization\\error on backgroud} & \tabincell{c}{ $5\%$ (appearance channels)\\$20\%$ (disappearance channels)}& $5\%$ (all channels) &$2\%$ (all channels)\\
\hline
\tabincell{c}{backgound sources} & \tabincell{c}{flavor misidentification;\\NC events;\\charge misidentification}& \tabincell{c}{flavor misidentification;\\NC events;\\intrinsic background} &\tabincell{c}{charge misidentification;\\NC events}\\
\hline
\tabincell{c}{channels} & \tabincell{c}{$\nu_e(\bar{\nu}_e)$ appearance\\$\nu_{\mu}(\bar{\nu}_{\mu})$ disappearance}& \tabincell{c}{$\nu_e(\bar{\nu}_e)$ appearance\\$\nu_{\mu}(\bar{\nu}_{\mu})$ disappearance} &\tabincell{c}{$\nu_{\mu}(\bar{\nu}_{\mu})$ appearance\\$\nu_{\mu}(\bar{\nu}_{\mu})$ disappearance\\$\nu_{e}(\bar{\nu}_{e})$ appearance}\\
\hline
\end{tabular}
\caption{Simulation details for T2HK, DUNE and the LENF in the current work.}
\label{tab:glbtable}
\end{table}

DUNE is a next-generation long baseline accelerator neutrino experiment with the primary goal of measuring 
the lepton CP-violating phase and the neutrino mass ordering.
In our simulation, we take a proton intensity power of 1.07 MW as the nominal neutrino beam configuration.
The energy range of the neutrino beam is from 500 MeV to 10 GeV, and the energy peak is located at around 3 GeV.
We consider a Liquid Argon Time-Projection Chamber (LArTPC) detector with 40 kton fiducial mass as the far detector 1300 km away from the beam source
and a near detector of 5 ton located at a baseline of 575 m.
For the energy resolution, we take the simulated migration matrices from the DUNE collaboration as our inputs.
Finally we assume seven years of the nominal running time with 3.5 years of the neutrino mode and 3.5 years of the antineutrino mode.
The T2HK experiment at Japan with a baseline of 295 km is an upgrade of the ongoing T2K experiment, which
will use a Water Cherenkov (WC) detector with a target mass 10 times larger than the existing Superkamiokande (SK) detector.
For our reference study a fiducial volume of 380 kton for two tanks (one tank for the first six years) with an energy resolution of $8.5\%/\sqrt{E}$ is considered here,
and the JPARC neutrino beamline is assumed to have a proton power of 1.3 MW from the beginning of T2HK.
As for the near detector a fiducial mass of 20 ton at the baseline 280 m is assumed.
Meanwhile, we assume ten years of data taking with six years of running for the first tank and four years of running for two tanks.
The ratio for the neutrino and antineutrino modes is one to three.
Regarding the LENF~\cite{FernandezMartinez:2010zza}, we assume the muon beam of 5 GeV with $10.6\times 10^{20}$ useful muon decays per year for each polarity.
We take a baseline of $L=1200$ km for the LENF and and consider a magnetized Totally Active Scintillator Detector (TASD)~\cite{FernandezMartinez:2010zza}
with the fiducial mass of 20 kton and an energy resolution of $15\%/\sqrt{E}$.
In addition, a near detector with the fiducial mass of 13 ton at the effective baseline of 2.28 km is assumed in our simulation~\cite{Tang:2009na}.

In the simulation, we employ the GLoBES software package~\cite{Huber:2004ka,Huber:2007ji} to simulate the aforementioned experiments.
The oscillation parameters of the three neutrino oscillation framework are take from Ref.~\cite{Esteban:2016qun},
which gives the fitting of oscillation parameters as $\Delta m^{2}_{21} = 7.50 \times 10^{-5}~{\rm eV}^2$, $\Delta m^{2}_{31} = 2.524 \times 10^{-3}~{\rm eV}^2$, $\theta^{}_{12} = 33.56^\circ$, $\theta^{}_{13} = 8.46^\circ$, $\theta^{}_{23} = 41.6^\circ$ and $\delta = 261^\circ$ (or $\Delta m^{2}_{21} = 7.50 \times 10^{-5}~{\rm eV}^2$, $\Delta m^{2}_{32} = 2.514 \times 10^{-3}~{\rm eV}^2$, $\theta^{}_{12} = 33.48^\circ$, $\theta^{}_{13} = 8.49^\circ$, $\theta^{}_{23} = 50^\circ$ and $\delta = 277^\circ$)
for the normal ordering case (or the inverted ordering case). We take the Earth crustal matter density as $3\,{\rm g}/{\rm cm}^3$ and the electron fraction to be 0.5 in the simulation. To have a clear illustration on the distinct DUV and IUV effects, our DUV framework defines with $\theta_{15} =\theta_{25}=\theta_{35}=0$, and the IUV scheme takes
$\theta_{14} =\theta_{24}=\theta_{34}=0$. In the general case, both $\theta_{i4}$ and $\theta_{i5}$ ($i=1,2,3$) will be arbitrary in the DUV$+$IUV framework.
To further reduce the number of parameters, the simplified scenarios with $\theta_{14}=\theta_{24}=\theta_{34}$ or $\theta_{15}=\theta_{25}=\theta_{35}$ are considered in our calculation. The values of the corresponding DUV and IUV parameters will be assigned in the next section when needed.

\section{Simulation results}

\begin{figure}
\includegraphics[scale=0.42]{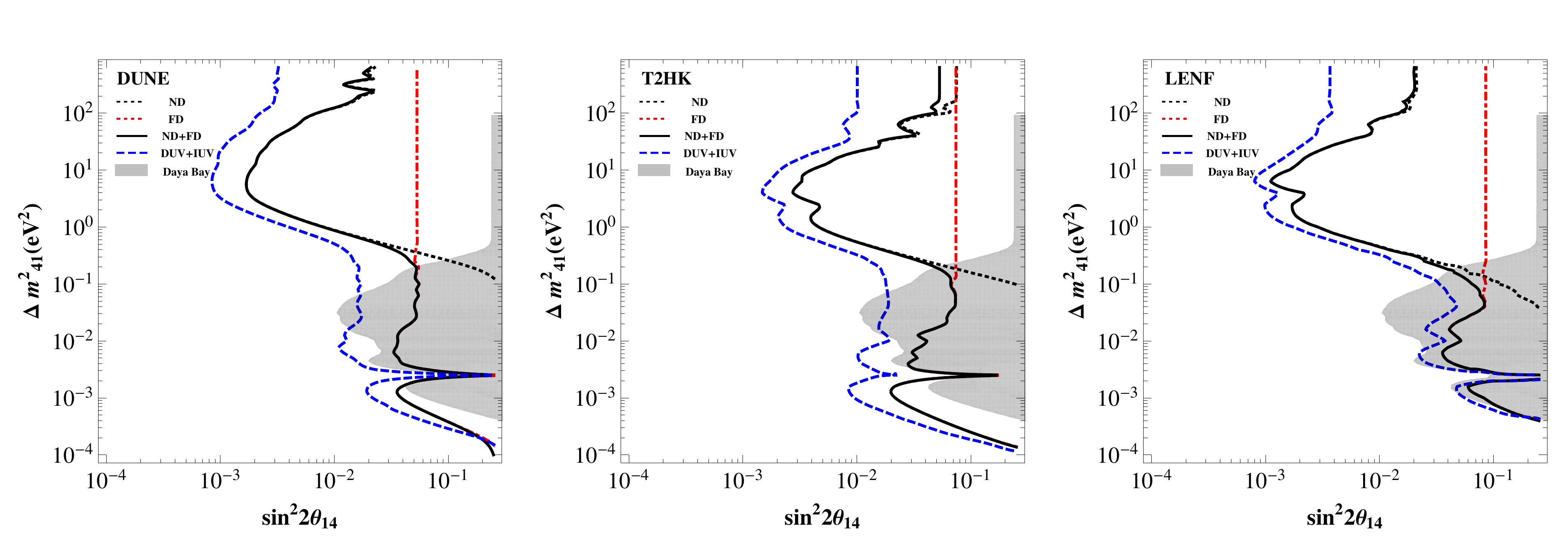}
\caption{
Exclusion curves in the $\sin^22\theta_{14} \textendash \Delta m^2_{41}$ plane at the confidence level (C.L.) of $2\sigma$ obtained from the near detectors (ND),
far detectors (FD) and their combinations of DUNE (left panel), T2HK (middle panel) and the LENF (right panel), respectively.
The latest exclusion limits from the Daya Bay reactor neutrino experiment~\cite{An:2016luf} is given as the grey shadowed region.
}
\label{lsn-sen}
\end{figure}
In this section, we present our simulation results of DUNE, T2HK and the LENF in the presence of the DUV and IUV effects.
{First, we start with a discussion on the potential to probe the DUV effects parametrized as a light sterile neutrino using both the near and far
detectors~\cite{Klop:2014ima,Palazzo:2015gja,Berryman:2015nua,Agarwalla:2016xxa,Choubey:2016fpi}
of the three experiments. To simulate, we generate the neutrino event spectra using the standard three neutrino oscillation paradigm.
When fitting the data, we carry out the analyses with two different hypotheses: the DUV case (the DUV parameters $\theta_{14}$ and
$\Delta m_{41}^2$ are scanned in their allowed regions) and the DUV$+$IUV case (the DUV parameters $\theta_{14}$ and $\Delta m_{41}^2$ are scanned in the allowed regions, and
the IUV parameters are fixed with $\theta_{i5}=5^\circ$ for $i=1,2,3$).}
{In Fig.~\ref{lsn-sen},
sensitivities to the DUV parameters obtained from the near (far) are given with the dashed black (red) curves. In addition, sensitivities to the DUV parameters are also shown with
(dashed blue lines) and without (solid black lines) the IUV effects using the combination of the near and far detectors.}
{Comparing the dashed black and red curves, one can observe that the oscillatory behavior of the DUV effect with $\Delta m_{41}^2 \gtrsim 0.1\;{\rm eV}^{2}$ only appears
at the near detector and will be averaged out at the far detector. Therefore, the far detector can only constrain the mixing angles $\theta_{i4}$ (i=1,2,3) rather than
the additional mass-squared difference, which will induce significant degeneracy between the DUV and IUV effects.
On the other hand, at the far detector, we can obtain another observable signal of spectral distortion for the smaller mass-squared difference with $\Delta m_{41}^2 \lesssim 0.1\;{\rm eV}^{2}$.} In order to illustrate the distinct features of DUV and IUV effects respectively,
we will take the additional mass-squared difference of the light sterile neutrino as $\Delta m^2_{41}\simeq 5\times10^{-3}$ eV$^2$.
{Furthermore, when we simultaneously consider both effects in the DUV$+$IUV framework, a stronger exclusion limit to $\theta_{14}$ is obtained compared to the DUV case.}

\begin{figure}
\includegraphics[scale=0.44]{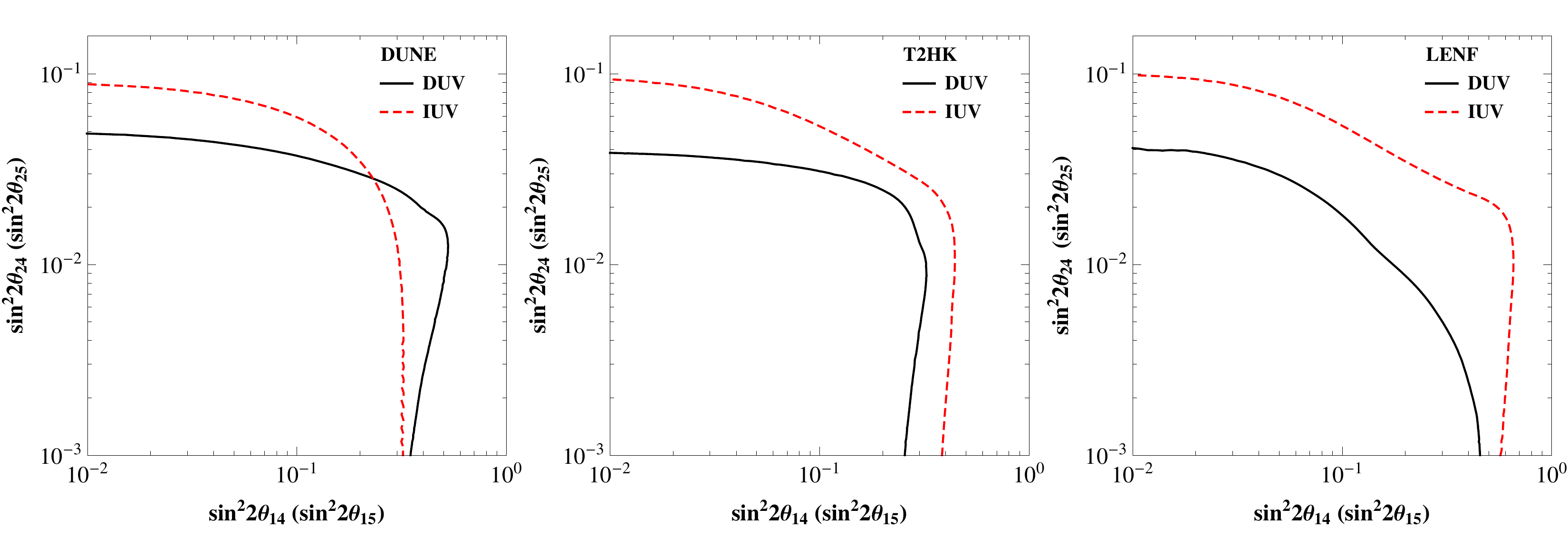}
\caption{Exclusion limits of $\sin^2 2\theta_{14} \textendash \sin^2 2\theta_{24}$ for DUV ($\sin^2 2\theta_{15} \textendash \sin^2 2\theta_{25}$ for IUV) obtained from DUNE (left panel), T2HK (middle panel) and the LENF (right panel). The C.L. is 2$\sigma$ for two degrees of freedom.}
\label{mixing-sen}
\end{figure}
After fixing the value of $\Delta m^2_{41}$, we are going to discuss the potential to constrain the active-sterile mixing of
the DUV and IUV effects at DUNE, T2HK and the LENF using their far detectors. In Fig.~\ref{mixing-sen}, we present the sensitivities to the mixing parameters
($\sin^22\theta_{14}$, $\sin^22\theta_{24}$) or ($\sin^22\theta_{15}$, $\sin^22\theta_{25}$) for the DUV (solid black lines) and IUV (dashed red lines) effects respectively.
In general, one can find that future accelerator facilities have better limits on the DUV mixing parameters rather than those of the IUV.
This is because light sterile neutrinos in the DUV scenario can induce both the rate and spectrum signatures in the far detector but only rate deficit can be observed for heavy sterile neutrinos in the IUV case.
The only exception is for the region of $\sin^22\theta_{14}> 0.3$ (or $\sin^22\theta_{15}> 0.3$) in DUNE. This property can be explained by the differences of normalization uncertainties in the experimental oscillation channels.
{In addition, one can observe that the limits to $\theta_{24}$ ($\theta_{25}$) are stronger than
those of $\theta_{14}$ ($\theta_{15}$). This can be illustrated by the oscillation probabilities.
In the appearance channels, $\theta_{14}$ and $\theta_{24}$ (or $\theta_{25}$ and $\theta_{25}$) contribute equally to the leading term of the neutrino transition probability.
However, in the $\nu_{\mu}$ and $\bar{\nu}_{\mu}$ disappearance channels, the leading contribution of the survival probability is from $\theta_{24}$ and $\theta_{25}$, while
$\theta_{14}$ and $\theta_{15}$ are left with marginal effects in the sub-leading terms. Regarding the unitarity violation test, we can obtain the levels of
0.06, 0.01 and 0.02 for the quantities of [$1-\sum^{3}_{i=1}|U_{e i}|^2$], [$1-\sum^{3}_{i=1}|U_{\mu i}|^2$] and [$\sum^{3}_{i=1}U_{ei}U^{\ast}_{\mu i}$] at the $2\sigma$ 
confidence level (C.L.) respectively. }
\begin{figure}
\includegraphics[scale=0.32]{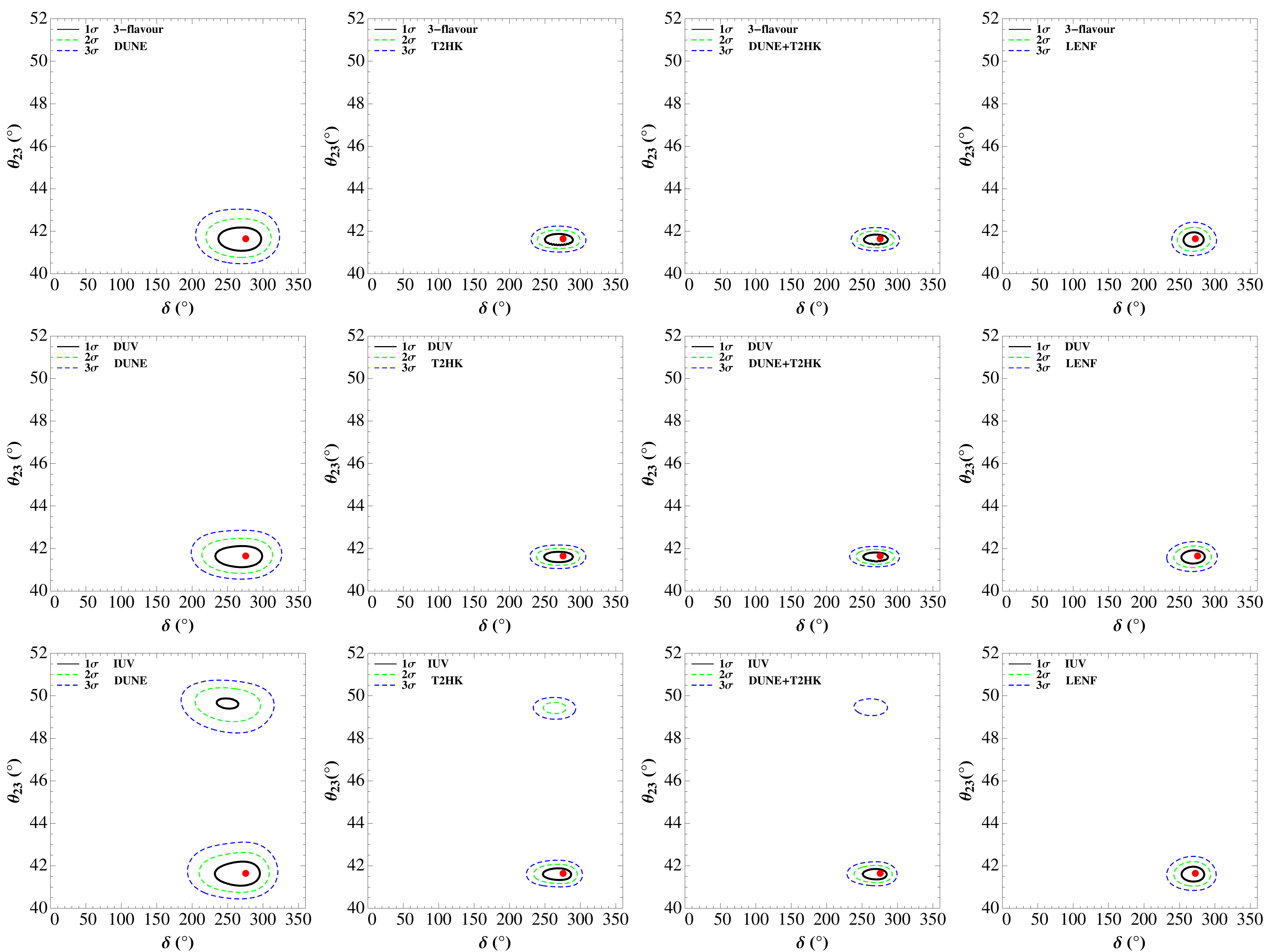}
\caption{Allowed region of the $\theta_{23} \textendash \delta_{}$ parameter space for DUNE only (first column), T2HK only (second column),
the combination of DUNE and T2HK (third column) and the LENF (fourth column).
The black, green and blue lines represent 1$\sigma$, 2$\sigma$ and 3$\sigma$ C.L., respectively. The upper, middle and lower panels represent the
frameworks of the standard three neutrino mixing, DUV and IUV, respectively.}
\label{fig:octantcp}
\end{figure}

Next we are going to discuss the DUV and IUV effects in the measurements of $\theta_{23}$ and the CP-violating phase $\delta$ in future accelerator neutrino facilities.
Current measurements of nearly maximal $\theta_{23}$ have been obtained in the atmospheric neutrino experiments~\cite{Kajita:2016vhj}
and long baseline accelerator neutrino experiments~\cite{Abe:2017bay,Adamson:2017gxd}. However, whether $\theta_{23}$ is smaller or larger than $45^\circ$ remains the
$\theta_{23}$ octant problem which is one of the main goal of future accelerator neutrino facilities.
The DUV and IUV effects may have significant effects on the measurement of the $\theta_{23}$ octant~\cite{Agarwalla:2016xlg,Agarwalla:2016fkh}.
Fig.~\ref{fig:octantcp} shows the allowed regions of $\theta_{23}$ and $\delta$ from the fits of the simulated data of DUNE only (first column), T2HK only (second column),
the combination of DUNE and T2HK (third column) and the LENF (fourth column), respectively. The upper panel illustrates the standard scenario of three-neutrino mixing,
and the middle and lower panels are shown for the cases of DUV and IUV frameworks.
The best-fit values are indicated with red points.
{In all panels the simulated data are generated with the standard three neutrino framework.
The upper, middle and lower panels illustrate the fitting results within the frameworks of the standard three neutrino mixing, DUV and IUV, respectively.
For the DUV and IUV frameworks, the additional mixing angles $\theta_{14}$ and $\theta_{15}$ are scanned in the range of [$0^\circ$, $10^\circ$] and the new CP-violating phase
$\delta_{new}$ changes in the range of [$0^\circ$, $360^\circ$].}

In the upper panel of the three neutrino mixing framework, DUNE, T2HK and LENF can determine the $\theta_{23}$ octant at more than {3$\sigma$ C.L.}, but
DUNE has larger uncertainties for $\theta_{23}$ and $\delta$ than T2HK and the LENF because of its relatively lower statistics.
On the precision of the oscillation parameters, T2HK can have the most accurate measurement of $\theta_{23}$ and the LENF has the best sensitivity on $\delta_{}$.
For the DUV scenario as shown in the middle panel of Fig.~\ref{fig:octantcp}, the presence of the light sterile neutrino can not destroy the
determination of the $\theta_{23}$ octant at the 3$\sigma$ C.L..
In the middle panel we can also observe that the precision of $\delta_{}$ will be reduced in the DUV scenario.
This is because additional CP-violating phases from active-sterile neutrino mixing can mimic the CP violation effect in the standard case.
In the lower panel of Fig.~\ref{fig:octantcp} we illustrate the IUV effects on the determination of the $\theta_{23}$ octant.
It is noted that the introduction of the IUV effect will induce degenerate solutions in the opposite octant region at the 1$\sigma$ C.L. for DUNE and 2$\sigma$ C.L. for T2HK.
However, the LENF is rather robust against the degenerate solution at better than 3$\sigma$ C.L..
The different effects of the DUV and IUV can be understood by the fact that IUV can only induce rate corrections to the three neutrino oscillation,
but DUV contributes both rate and spectrum signatures to the experimental measurements. Note that our DUV effect on the $\theta_{23}$ octant is different from that in
Refs.~\cite{Agarwalla:2016xlg,Agarwalla:2016fkh} in the presence of the eV-scale sterile neutrino. The relatively larger mass-squared difference $\Delta m^2_{41}$
in Ref.~\cite{Agarwalla:2016xlg,Agarwalla:2016fkh} will be averaged out in the far detectors.
Therefore, their conclusion is rather similar to our IUV case with the severe degeneracy problem in the $\theta_{23}$ octant determination.

\begin{figure}
\includegraphics[scale=0.3]{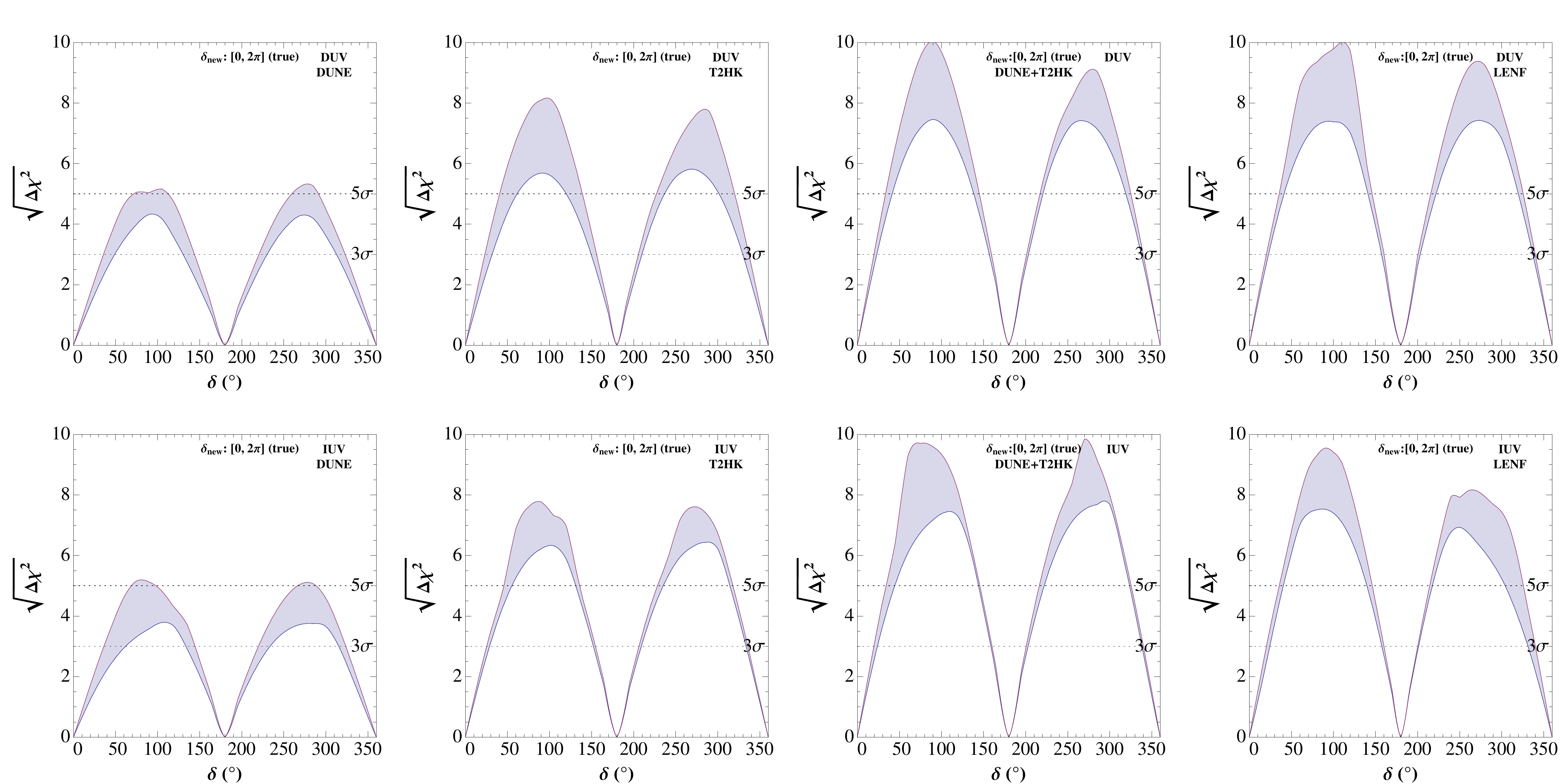}
\caption{
Discovery reach of the standard-like CP-violating phase $\delta$ for DUNE only (first column), T2HK only (second column),
the combination of DUNE and T2HK (third column) and the LENF (fourth column). The upper and lower panels represent the
frameworks of the DUV and IUV, respectively.}
\label{fig:CP1}
\end{figure}

\begin{figure}
\includegraphics[scale=0.3]{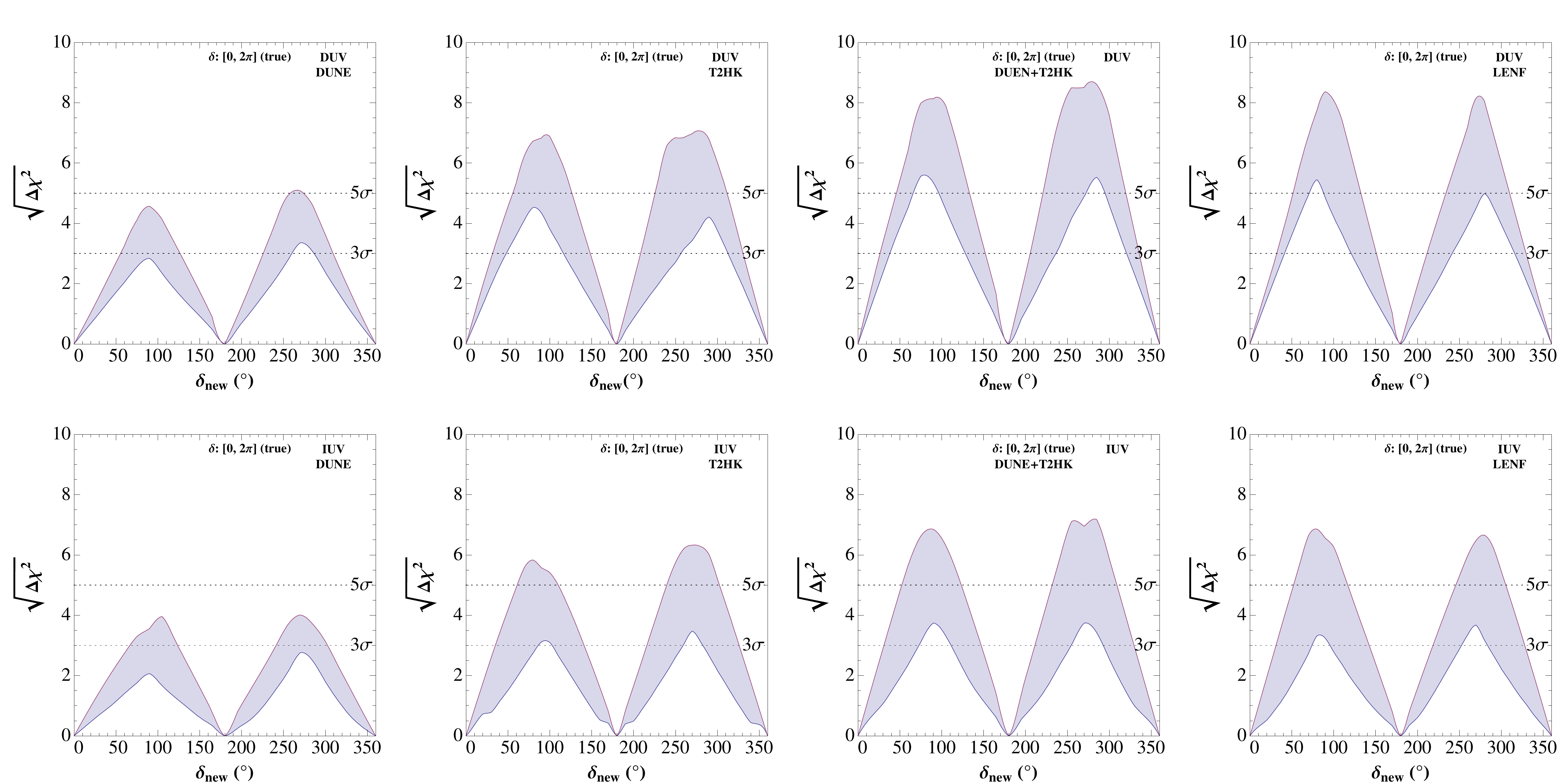}
\caption{
Discovery reach of the effective new CP-violating phase $\delta_{new}$ for DUNE only (first column), T2HK only (second column),
the combination of DUNE and T2HK (third column) and the LENF (fourth column). The upper and lower panels represent the
frameworks of the DUV and IUV, respectively.}
\label{fig:CP2}
\end{figure}

The DUV and IUV effects may have significant impact on the CP violation measurements~\cite{Forero:2016cmb,Miranda:2016wdr,Ge:2016xya,Escrihuela:2016ube,Kelly:2017kch,Rout:2017udo}.
In Fig.~\ref{fig:CP1} we show the discovery reach of the standard-like CP-violating phase $\delta$ for DUNE only (first column), T2HK only (second column),
the combination of DUNE and T2HK (third column) and the LENF (fourth column). The upper and lower panels represent the frameworks of the DUV and IUV, respectively.
Here $\Delta \chi^2=|\chi^2(\delta)-\chi^2(0^\circ\; {\rm or} \;180^\circ)|$ is defined as the absolute difference between the $\chi^2$ function at varying true values of the
CP-violating phase $\delta$ and $\chi^2$ at $\delta=0^{\circ}/180^\circ$.
In the DUV (IUV) case, we have taken the mixing angles as $\theta_{14}=\theta_{24}=\theta_{34}
=10^\circ$ ($\theta_{15}=\theta_{25}=\theta_{35}=10^\circ$).
Since the additional CP-violating phases $\delta_{i4}$ ($\delta_{i5}$) would contribute to the neutrino oscillation behavior in a sophisticated way, 
for simplicity, we only consider $\delta_{24}$ ($\delta_{25}$) as the effective new CP violating phase $\delta_{new}$ and take the other two phases to be zero.
The variation of $\delta_{new}$ gives rise to the blue bands in Fig.~\ref{fig:CP1}.
In a similar way, Fig.~\ref{fig:CP2} shows the discovery reach of CP violation induced by the effective new CP-violating phase $\delta_{new}$.
The variation of the standard CP-violating phase $\delta$ is shown with the blue bands. All the other undisplayed parameters are marginalized.

Fig.~\ref{fig:CP1} shows that the discovery potential of standard CP violation taking into account the DUV and IUV effects would be highly suppressed compared to
the standard three neutrino framework in all three experiments.
In particular, the discovery potential of maximal CP violation caused by $\delta_{}$ would be
degraded to the levels of $4.2\sigma$ ($3.7\sigma$), $5.6\sigma$ ($6.2\sigma$) and $7.4\sigma$ ($6.5\sigma$) at DUNE, T2HK, and the LENF respectively for the DUV (IUV) case.
As for the results of Fig.~\ref{fig:CP2}, the discovery potential of maximal CP violation induced by $\delta_{new}$ would be
reduced to 2.8$\sigma$ ($2.1\sigma$), 3.7$\sigma$ ($3.1\sigma$) and 5.0$\sigma$ ($3.2\sigma$) at DUNE, T2HK, and the LENF respectively for the DUV (IUV) case.
In the respect, the LENF has the best performance of measuring the CP-violating phases.
We can observe that the LENF can still have an excellent discovery reach of maximal
CP violation with the sensitivities better than 6$\sigma$ for $\delta_{}$ induced maximal CP violation
and better than 3$\sigma$ for $\delta_{new}$ induced maximal CP violation in the presence of DUV and IUV effects.
Comparing the simulation results between Fig.~\ref{fig:CP1}
and Fig.~\ref{fig:CP2}, we can observe that the discovery potential of CP violation induced by $\delta$ is better than that induced by $\delta_{new}$. This is because the standard
CP-violating phase $\delta$ have more significant contribution than that of the new CP-violating phase as illustrated in the perturbative expansion
of the neutrino oscillation probabilities given in Ref.~\cite{Li:2015oal}.

Because of the complementary roles of accelerator neutrino experiments with distinct baseline, beam configurations
and neutrino oscillation channels~\cite{Coloma:2012ji,Cao:2015ita,Fukasawa:2016yue,Ballett:2016daj},
it is powerful to combine DUNE of the wide-band neutrino beam and long baseline with T2HK of the off-axis narrow neutrino beam and medium baseline.
The synergistic effects of the DUNE and T2HK combination on the $\theta_{23}$ octant and CP violation in the presence of DUV and IUV effects are illustrated
in the third columns of Fig.~\ref{fig:octantcp}, Fig.~\ref{fig:CP1} and Fig.~\ref{fig:CP2}. We can find from the third column of Fig.~\ref{fig:octantcp}
that the degenerate solution of $\theta_{23}$ in the opposite octant region appears at the 3$\sigma$ C.L., and the precision of oscillation parameters is also improved.
From the third columns of Fig.~\ref{fig:CP1} and Fig.~\ref{fig:CP2}, the discovery potential of maximal CP violation caused by $\delta_{}$ ($\delta_{new}$) would be
$7.6\sigma$ ($5.5\sigma$) for the DUV case and $7.2\sigma$ ($3.7\sigma$) for the IUV case. This observation demonstrates that the sensitivity of
the DUNE and T2HK combination is comparative to that of the LENF in the CP violation measurement.

Based on the above simulation results, we briefly summarize the feature of the impact from DUV and IUV effects.
First, at the oscillation probability level, the DUV is parametrized by an additional mass-squared difference $\Delta m_{41}^2$ and the new mixing angles,
which will lead to the observable effects with both the rate and spectrum information. On the other hand, the IUV only has new mixing parameters
which will only affect the oscillation amplitudes in the probability. In this regard,
the new mixing angles and the standard neutrino mixing angles (e.g., $\theta_{23}$) may have the same effects and will result in
possible degenerate solutions to the mixing angle $\theta_{23}$.
Second, both the standard CP-violating phase (i.e., $\delta$) and the new CP-violating phase (i.e., $\delta_{new}$) can induce observable CP violation in
the appearance channels of neutrino oscillations. Therefore, the variation of one particular CP-violating phase in the parameter space would induce ambiguity to
the other one, and thus reduce the sensitivity of the CP violation.
Finally, we stress that an effective way to disentangle the effect of the standard three neutrino framework from the DUV and IUV effects
is to combine different experiments with distinct baseline, beam configurations and neutrino oscillation channels.

\section{Summary}
{New physics beyond the Standard Model could cause new effects on the standard three neutrino oscillation paradigm. In this work, we have investigated the impact of one light sterile
neutrino (i.e., direct unitarity violation) and one heavy sterile neutrino (i.e., indirect unitarity violation) in future accelerator neutrino facilities, DUNE, T2HK and the LENF. We have obtained the exclusion limits in the $\sin^2\theta_{14}-\Delta m_{41}^2$ plane of the DUV scheme with both near and far detectors.
The near detectors are sensitive to the additional mass-squared difference with $\Delta m_{41}^2 \gtrsim 0.1\;{\rm eV}^{2}$. In addition, with the help of the far detectors,
we can extend the exclusion limits to the smaller mass-squared difference region of $\Delta m_{41}^2 \lesssim 0.1\;{\rm eV}^{2}$.
However, the IUV effect contributes to the amplitudes of oscillation terms and thus is only sensitive to the rate information in both the near and far detectors.
Regarding the test of leptonic unitarity violation, we can obtain the levels of
0.06, 0.01 and 0.02 for the quantities of [$1-\sum^{3}_{i=1}|U_{e i}|^2$], [$1-\sum^{3}_{i=1}|U_{\mu i}|^2$] and [$\sum^{3}_{i=1}U_{ei}U^{\ast}_{\mu i}$] at the $2\sigma$ C.L. respectively.

{Both the DUV and IUV effects, if unknown, may induce biased solutions for the three neutrino mixing parameters,
such as the octant of $\theta_{23}$ and the CP-violating phase $\delta_{}$.
It is observed that future accelerator facilities are rather robust against the DUV effect in the $\theta_{23}$ measurement,
and the presence of the light sterile neutrino can not destroy the determination of the $\theta_{23}$ octant at the {3$\sigma$ C.L. for all the three experiments}.
However, the IUV effect will be critical for the determination of the $\theta_{23}$ octant.
It is noted that the introduction of the IUV effect will induce degenerate solutions in the opposite octant region at the 1$\sigma$ C.L. for DUNE and 2$\sigma$ C.L. for T2HK.
However, the LENF can determine the octant of $\theta_{23}$ better than the 3$\sigma$ C.L..
As for the discovery reach of the CP violation, because of the ambiguity of multiple sources of leptonic CP violation from $\delta$ and $\delta_{new}$,
the discovery reach of CP violation from one particular source of $\delta_{}$ or $\delta_{new}$ would be significantly degraded by the other one.
In the aspect, the LENF has the best performance of measuring the CP-violating phases with the sensitivities better than 6$\sigma$ for $\delta_{}$ induced maximal CP violation
and better than 3$\sigma$ for $\delta_{new}$ induced maximal CP violation in the presence of DUV and IUV effects.
Finally, we want to stress that the combination of experiments with different oscillation channels, different neutrino
beams and different detector techniques will be an effective solution to the parameter degeneracy problem and give the robust measurement of the $\theta_{23}$ octant and the leptonic CP violation even if the direct and indirect unitarity violation are taken into account. We hope future accelerator facilities could help us to pin down the standard
three neutrino paradigm and search for new physics beyond the Standard Model.}

%
%

\section*{Acknowledgements}

This work was in part supported by the National Natural Science Foundation of China under Grant Nos.~11505301, 11305193 and 11135009,
by the Strategic Priority Research Program of the Chinese Academy of Sciences under Grant No.~XDA10010100, by the Special Program 
for Applied Research on Super Computation of the NSFC-Guangdong Joint Fund (the second phase) under Grant No.~U1501501, 
and by the CAS Center for Excellence in Particle Physics (CCEPP). YBZ appreciated useful discussions with Steven Wong.


\begin{thebibliography}{99}

\bibitem{Kajita:2016cak}
  T.~Kajita,
  Rev.\ Mod.\ Phys.\  {\bf 88}, 030501 (2016).

\bibitem{McDonald:2016ixn}
  A.~B.~McDonald,
  Rev.\ Mod.\ Phys.\  {\bf 88}, 030502 (2016).

\bibitem{Olive:2016xmw}
  C.~Patrignani {\it et al.} [Particle Data Group],
  Chin.\ Phys.\ C {\bf 40}, 100001 (2016).

\bibitem{Pontecorvo:1957cp}
  B.~Pontecorvo,
  Sov.\ Phys.\ JETP {\bf 6}, 429 (1957)
  [Zh.\ Eksp.\ Teor.\ Fiz.\  {\bf 33}, 549 (1957)].

\bibitem{Maki:1962mu}
  Z.~Maki, M.~Nakagawa and S.~Sakata,
  Prog.\ Theor.\ Phys.\  {\bf 28}, 870 (1962).

\bibitem{sterile}
  K.~N.~Abazajian, M.~A.~Acero, S.~K.~Agarwalla, A.~A.~Aguilar-Arevalo, C.~H.~Albright {\it et al.},
  arXiv:1204.5379 [hep-ph];

  \bibitem{Gariazzo:2017fdh}
  S.~Gariazzo, C.~Giunti, M.~Laveder and Y.~F.~Li,
  JHEP {\bf 1706}, 135 (2017)
  [arXiv:1703.00860 [hep-ph]].

\bibitem{Gariazzo:2015rra}
  S.~Gariazzo, C.~Giunti, M.~Laveder, Y.~F.~Li and E.~M.~Zavanin,
  J.\ Phys.\ G {\bf 43}, (2016) 033001
  [arXiv:1507.08204 [hep-ph]].

\bibitem{Giunti:2013aea}
  C.~Giunti, M.~Laveder, Y.~F.~Li and H.~W.~Long,
  Phys.\ Rev.\ D {\bf 88}, 073008 (2013)
  [arXiv:1308.5288 [hep-ph]].

\bibitem{Kopp:2013vaa}
  J.~Kopp, P.~A.~N.~Machado, M.~Maltoni and T.~Schwetz,
  JHEP {\bf 1305}, 050 (2013)
  [arXiv:1303.3011 [hep-ph]];

\bibitem{Adhikari:2016bei}
  M.~Drewes {\it et al.},
  JCAP {\bf 1701}, 025 (2017)
  [arXiv:1602.04816 [hep-ph]].

\bibitem{Kusenko:2009up}
  A.~Kusenko,
  Phys.\ Rept.\  {\bf 481}, 1 (2009)
  [arXiv:0906.2968 [hep-ph]];

\bibitem{Boyarsky:2009ix}
  A.~Boyarsky, O.~Ruchayskiy and M.~Shaposhnikov,
  Ann.\ Rev.\ Nucl.\ Part.\ Sci.\  {\bf 59}, 191 (2009)
  [arXiv:0901.0011 [hep-ph]];

\bibitem{Araki:2011zg}
  T.~Araki and Y.~F.~Li,
  Phys.\ Rev.\ D {\bf 85}, 065016 (2012)
  [arXiv:1112.5819 [hep-ph]];

\bibitem{Minkowski:1977sc}
  P.~Minkowski,
  Phys.\ Lett.\  {\bf 67B}, 421 (1977).

\bibitem{Yanagida:1979ss}
T. Yanagida, In {\it Proceedings of the Workshop on Unified Theory and
the Baryon Number of the Universe}, edited by O. Sawada and A.
Sugamoto, (KEK, Tsukuba, 1979), p. 95.

\bibitem{Gell-Mann:1979ss}
M. Gell-Mann, P. Ramond and R. Slansky, In {\it Supergravity},
edited by P. van Nieuwenhuizen and D. Z. Freeman, (North-Holland,
Amsterdam, 1979), p. 315.

\bibitem{Glashow:1979ss}
S. L. Glashow, In {\it Quarks and Leptons}, edited by M. Levy {\it
et al.} (Plenum, New York, 1980), p. 707.

\bibitem{Mohapatra:1979ia}
  R.~N.~Mohapatra and G.~Senjanovic,
  Phys.\ Rev.\ Lett.\  {\bf 44}, 912 (1980).

\bibitem{Antusch:2006vwa}
  S.~Antusch, C.~Biggio, E.~Fernandez-Martinez, M.~B.~Gavela and J.~Lopez-Pavon,
  JHEP {\bf 0610}, 084 (2006)
  [hep-ph/0607020].

\bibitem{Xing:2012kh}
  Z.~Z.~Xing,
  Phys.\ Lett.\ B {\bf 718}, 1447 (2013)
  [arXiv:1210.1523 [hep-ph]].

\bibitem{Qian:2013ora}
  X.~Qian, C.~Zhang, M.~Diwan and P.~Vogel,
  arXiv:1308.5700 [hep-ex].

\bibitem{Luo:2014fia}
  S.~Luo,
  Int.\ J.\ Mod.\ Phys.\ A {\bf 29}, 1444006 (2014).

\bibitem{Antusch:2014woa}
  S.~Antusch and O.~Fischer,
  JHEP {\bf 1410}, 94 (2014)
  [arXiv:1407.6607 [hep-ph]].

\bibitem{Escrihuela:2015wra}
  F.~J.~Escrihuela, D.~V.~Forero, O.~G.~Miranda, M.~Tortola and J.~W.~F.~Valle,
  Phys.\ Rev.\ D {\bf 92}, 053009 (2015)
  [arXiv:1503.08879 [hep-ph]].

\bibitem{Li:2015oal}
  Y.~F.~Li and S.~Luo,
  Phys.\ Rev.\ D {\bf 93}, 033008 (2016)
  [arXiv:1508.00052 [hep-ph]].

\bibitem{Parke:2015goa}
  S.~Parke and M.~Ross-Lonergan,
  Phys.\ Rev.\ D {\bf 93}, 113009 (2016)
  [arXiv:1508.05095 [hep-ph]].

\bibitem{Fong:2016yyh}
  C.~S.~Fong, H.~Minakata and H.~Nunokawa,
  JHEP {\bf 1702}, 114 (2017)
  [arXiv:1609.08623 [hep-ph]].

\bibitem{Blennow:2016jkn}
  M.~Blennow, P.~Coloma, E.~Fernandez-Martinez, J.~Hernandez-Garcia and J.~Lopez-Pavon,
  JHEP {\bf 1704}, 153 (2017)
  [arXiv:1609.08637 [hep-ph]].

\bibitem{Acciarri:2015uup}
  R.~Acciarri {\it et al.} [DUNE Collaboration],
  arXiv:1512.06148 [physics.ins-det].

\bibitem{Abe:2014oxa}
  K.~Abe {\it et al.} [Hyper-Kamiokande Working Group],
  arXiv:1412.4673 [physics.ins-det].

\bibitem{FernandezMartinez:2010zza}
  E.~Fernandez Martinez, T.~Li, S.~Pascoli and O.~Mena,
  Phys.\ Rev.\ D {\bf 81}, 073010 (2010)
  [arXiv:0911.3776 [hep-ph]].

\bibitem{Huber:2004ka}
  P.~Huber, M.~Lindner and W.~Winter,
  Comput.\ Phys.\ Commun.\  {\bf 167}, 195 (2005)
  [hep-ph/0407333].

\bibitem{Huber:2007ji}
  P.~Huber, J.~Kopp, M.~Lindner, M.~Rolinec and W.~Winter,
  Comput.\ Phys.\ Commun.\  {\bf 177}, 432 (2007)
  [hep-ph/0701187].

  \bibitem{Xing2012}
  Z.~Z.~Xing,
  Phys.\ Rev.\ D {\bf 85}, 013008 (2012)
  [arXiv:1110.0083 [hep-ph]].

\bibitem{Dev:2012bd}
  P.~S.~Bhupal Dev and A.~Pilaftsis,
  Phys.\ Rev.\ D {\bf 87}, 053007 (2013)
  [arXiv:1212.3808 [hep-ph]].

\bibitem{Bilenky:1980cx}
  S.~M.~Bilenky, J.~Hosek and S.~T.~Petcov,
  Phys.\ Lett.\ B {\bf 94}, 495 (1980).

\bibitem{Langacker:1988up}
  P.~Langacker and D.~London,
  Phys.\ Rev.\ D {\bf 38}, 907 (1988).

\bibitem{Kopp:2007ne}
  J.~Kopp, M.~Lindner, T.~Ota and J.~Sato,
  Phys.\ Rev.\ D {\bf 77}, 013007 (2008).

\bibitem{Agarwalla:2014bsa}
  S.~K.~Agarwalla, P.~Bagchi, D.~V.~Forero and M.~T¨®rtola,
  JHEP {\bf 1507}, 060 (2015).

\bibitem{Tang:2009na}
  J.~Tang and W.~Winter,
  Phys.\ Rev.\ D {\bf 80}, 053001 (2009)
  [arXiv:0903.3039 [hep-ph]].

\bibitem{Esteban:2016qun}
  I.~Esteban, M.~C.~Gonzalez-Garcia, M.~Maltoni, I.~Martinez-Soler and T.~Schwetz,
  JHEP {\bf 1701}, 087 (2017)
  [arXiv:1611.01514 [hep-ph]].

\bibitem{Klop:2014ima}
  N.~Klop and A.~Palazzo,
  Phys.\ Rev.\ D {\bf 91}, 073017 (2015)
  [arXiv:1412.7524 [hep-ph]].

\bibitem{Palazzo:2015gja}
  A.~Palazzo,
  Phys.\ Lett.\ B {\bf 757}, 142 (2016)
  [arXiv:1509.03148 [hep-ph]].

\bibitem{Berryman:2015nua}
  J.~M.~Berryman, A.~de Gouv¨ºa, K.~J.~Kelly and A.~Kobach,
  Phys.\ Rev.\ D {\bf 92}, no. 7, 073012 (2015)
  [arXiv:1507.03986 [hep-ph]].

\bibitem{Agarwalla:2016xxa}
  S.~K.~Agarwalla, S.~S.~Chatterjee and A.~Palazzo,
  JHEP {\bf 1609}, 016 (2016)
  [arXiv:1603.03759 [hep-ph]].

\bibitem{Choubey:2016fpi}
  S.~Choubey and D.~Pramanik,
  Phys.\ Lett.\ B {\bf 764}, 135 (2017)
  [arXiv:1604.04731 [hep-ph]].

\bibitem{An:2016luf}
  F.~P.~An {\it et al.} [Daya Bay Collaboration],
  Phys.\ Rev.\ Lett.\  {\bf 117}, 151802 (2016)
  [arXiv:1607.01174 [hep-ex]].

\bibitem{Kajita:2016vhj}
  T.~Kajita {\it et al.} [Super-Kamiokande Collaboration],
  Nucl.\ Phys.\ B {\bf 908}, 14 (2016).

\bibitem{Abe:2017bay}
  K.~Abe {\it et al.} [T2K Collaboration],
  Phys.\ Rev.\ D {\bf 96}, 011102 (2017)
  [arXiv:1704.06409 [hep-ex]].

\bibitem{Adamson:2017gxd}
  P.~Adamson {\it et al.} [NOvA Collaboration],
  Phys.\ Rev.\ Lett.\  {\bf 118}, 231801 (2017)
  [arXiv:1703.03328 [hep-ex]].

  \bibitem{Agarwalla:2016xlg}
  S.~K.~Agarwalla, S.~S.~Chatterjee and A.~Palazzo,
  Phys.\ Rev.\ Lett.\  {\bf 118}, 031804 (2017)
  [arXiv:1605.04299 [hep-ph]].

\bibitem{Agarwalla:2016fkh}
  S.~K.~Agarwalla, S.~S.~Chatterjee and A.~Palazzo,
  Phys.\ Lett.\ B {\bf 762}, 64 (2016)
  [arXiv:1607.01745 [hep-ph]].

\bibitem{Forero:2016cmb}
  D.~V.~Forero and P.~Huber,
  Phys.\ Rev.\ Lett.\  {\bf 117}, 031801 (2016)
  [arXiv:1601.03736 [hep-ph]].

\bibitem{Miranda:2016wdr}
  O.~G.~Miranda, M.~Tortola and J.~W.~F.~Valle,
  Phys.\ Rev.\ Lett.\  {\bf 117}, 061804 (2016)
  [arXiv:1604.05690 [hep-ph]].
  
\bibitem{Ge:2016xya}
  S.~F.~Ge, P.~Pasquini, M.~Tortola and J.~W.~F.~Valle,
  Phys.\ Rev.\ D {\bf 95}, no. 3, 033005 (2017)
  [arXiv:1605.01670 [hep-ph]].

\bibitem{Escrihuela:2016ube}
  F.~J.~Escrihuela, D.~V.~Forero, O.~G.~Miranda, M.~T¨®rtola and J.~W.~F.~Valle,
  arXiv:1612.07377 [hep-ph].

\bibitem{Kelly:2017kch}
  K.~J.~Kelly,
  Phys.\ Rev.\ D {\bf 95}, 115009 (2017)
  [arXiv:1703.00448 [hep-ph]].

\bibitem{Rout:2017udo}
  J.~Rout, M.~Masud and P.~Mehta,
  Phys.\ Rev.\ D {\bf 95}, 075035 (2017)
  [arXiv:1702.02163 [hep-ph]].

\bibitem{Coloma:2012ji}
  P.~Coloma, P.~Huber, J.~Kopp and W.~Winter,
  Phys.\ Rev.\ D {\bf 87}, 033004 (2013)
  [arXiv:1209.5973 [hep-ph]].

\bibitem{Cao:2015ita}
  J.~Cao {\it et al.} [ICFA Neutrino Panel Collaboration],
  arXiv:1501.03918 [physics.acc-ph].

\bibitem{Fukasawa:2016yue}
  S.~Fukasawa, M.~Ghosh and O.~Yasuda,
  Nucl.\ Phys.\ B {\bf 918}, 337 (2017)
  [arXiv:1607.03758 [hep-ph]].

\bibitem{Ballett:2016daj}
  P.~Ballett, S.~F.~King, S.~Pascoli, N.~W.~Prouse and T.~Wang,
  arXiv:1612.07275 [hep-ph].

\end{thebibliography}
\end{document}